\documentclass[prl,twocolumn,showpacs,superscriptaddress,floatfix,reprint]{revtex4}
\usepackage[utf8]{inputenc}
\usepackage[english]{babel}
\usepackage[T1]{fontenc}
\usepackage{graphicx}
\usepackage{amsmath}
\usepackage{amsfonts}
\usepackage{amssymb}
\usepackage{siunitx}
\usepackage{textcomp}
\usepackage{standalone}
\usepackage{xspace}
\usepackage[caption=false]{subfig}
\usepackage[normalem]{ulem}
\usepackage{color}
\begin{document}
\title{Foam as a self-assembling amorphous photonic bandgap material}

% Use letters for affiliations, numbers to show equal authorship (if applicable) and to indicate the corresponding author
\author{Joshua Ricouvier}
\affiliation{ESPCI Paris, PSL Research University, CNRS, IPGG, MMN, 6 rue Jean Calvin, F-75005, Paris, France}
\author{Patrick Tabeling}
\affiliation{ESPCI Paris, PSL Research University, CNRS, IPGG, MMN, 6 rue Jean Calvin, F-75005, Paris, France}
\author{Pavel Yazhgur}
\affiliation{ESPCI Paris, PSL Research University, CNRS, IPGG, MMN, 6 rue Jean Calvin, F-75005, Paris, France}

\begin{abstract}
We show that slightly polydisperse disordered two-dimensional foams can be used as a self-assembled template for isotropic photonic bandgap (PBG) materials for TE polarization. Calculations based on in-house experimental and simulated foam structures demonstrate that, at sufficient refractive index contrast, a dry foam organization with 3-fold nodes and long slender Plateau borders is especially advantageous to open a large PBG. A transition from dry to wet foam structure rapidly closes the PBG mainly by formation of  bigger 4-fold nodes filling PBG with defect modes. By tuning the foam area fraction, we find an optimal quantity of dielectric material which maximises the PBG in experimental systems. The obtained results have a potential to be extended to 3D foams in order to produce a next generation of self-assembled disordered PBG materials enabling fabrication of cheap and scalable photonic devices.
\end{abstract}

%\dates{This manuscript was compiled on \today}
%\doi{\url{www.pnas.org/cgi/doi/10.1073/pnas.XXXXXXXXXX}}

\maketitle
%\thispagestyle{firststyle}
%\ifthenelse{\boolean{shortarticle}}{\ifthenelse{\boolean{singlecolumn}}{\abscontentformatted}{\abscontent}}{}

% If your first paragraph (i.e. with the \dropcap) contains a list environment (quote, quotation, theorem, definition, enumerate, itemize...), the line after the list may have some extra indentation. If this is the case, add \parshape=0 to the end of the list environment.

\section{Introduction}

\begin{figure*}[t!]
\begin{center}
\includegraphics[width=1\linewidth]{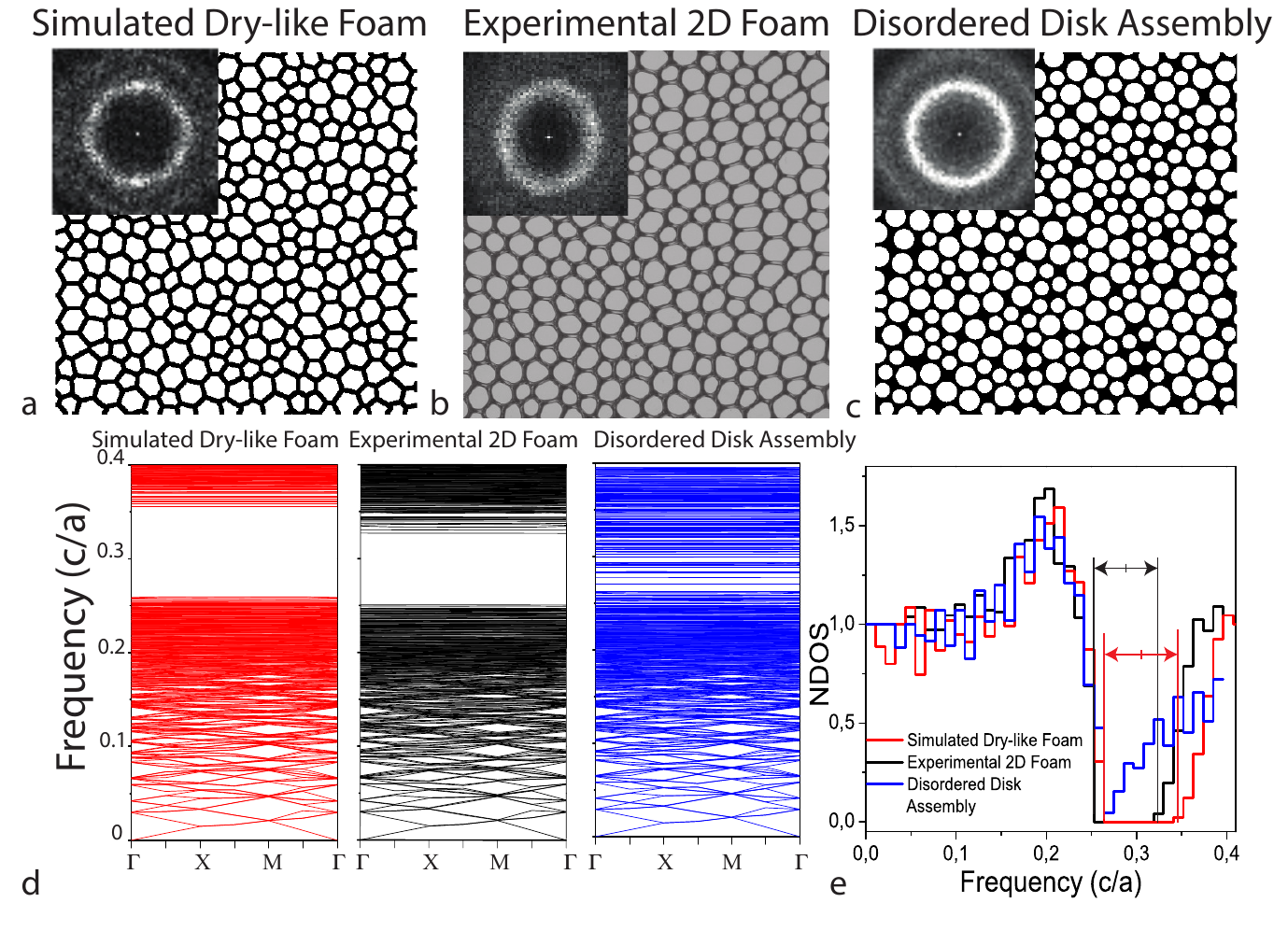}
\caption{\textbf{(a)} Binary image of a simulated dry-like foam generated from a random close pack configuration followed by Voronoi tessellation and annealing with Surface Evolver software. Top-Left: Close up on the Fourier transform of the image. \textbf{(b)} Photograph of a experimental 2D foam generated in a Hele-Shaw cell. Top-Left: Close up on the Fourier transform of the image. \textbf{(c)} Disordered disk assembly representing very wet foam. Top-Left: Close up on the Fourier transform of the image. For all three systems, ratio of bubble sizes is about 0.75, while the fraction of big bubbles is about 0.55. The area fraction of the continuous phase is  about 0.40. \textbf{(d)} Corresponding band diagrams. We set the refractive index contrast to 3.4. The frequencies are normalised with characteristic length defined as $a=L/\sqrt{N_{bubbles}}$, $L$ being the dimension of the image and $N_{bubbles}$ the number of bubbles inside the image \textbf{(e)} Normalized density of optical states for each system. Band gap widths are  $28\%$ and $22\%$ for the simulated dry-like foam and the experimental 2D foam, respectively. Only a pseudogap can be observed for disordered disk assemblies.}
\label{Imagesoffoam}
\end{center}
\end{figure*}

Materials with a complete photonic bandgap (PBG), which prevent the propagation of electromagnetic waves in all directions within a certain frequency range, have given rise to various promising industrial applications, such as lossless wave guides \cite{mekis1996high,ishizaki2013realization}, LEDs with high light extraction efficiency \cite{david2012}, high-Q laser cavities \cite{altug2006,amoah2015high} and optical elements of computers \cite{nozaki2012}. The search for PBG materials first led the scientific community to consider long range ordered structures, such as photonic crystals and quasicrystals, where formation of PBG is usually related to the multiple coherent Bragg scattering \cite{yablonovitch1987inhibited,joannopoulos2011photonic,man2005experimental}. However, fabrication issues and high sensitivity to defects strongly limit the development of potential devices based on photonic crystals \cite{vlasov2001chip, li2000fragility, li2001photonic}. Recently discovered disordered materials with a large complete PBG are argued to be potentially easier to fabricate \cite{florescu2009designer}. Moreover, such disordered photonic materials also offer directional isotropy, useful to create free form wave guides, isotropic radiation sources or non-iridescent structural color pigments \cite{florescu2013optical,florescu2009designer,lopez2018true} - applications simply impossible for classical photonic crystals. Disordered structures possessing PBG are interesting for the study of different regimes of optical transport such as the Anderson localization \cite{froufe2017band,wiersma2013disordered}.

There exist different disordered structures possessing a wide isotropic bandgap, which share some peculiar features. Champions of disordered PBG networks are usually constant valency connected networks of dielectric material surrounded by air, for example, 4-fold amorphous diamond or 3-fold gyroid structures \cite{edagawa2008photonic,wu2018optical,  li2018biological}. Similar behavior is observed in 3-fold coordinated 2D network systems for TE polarization (electric field in the plane) which is often used to predict properties of real 3D systems with decreased computational cost \cite{sellers2017local}. Within the same class (e.g., valency) of structures, different design protocols have been applied to find disordered networks with the largest possible bandgaps. One of the first optimization techniques derives networks from stealthy hyperuniform point patterns \cite{florescu2009designer} exhibiting zero structure factor for wave vectors below a certain limit\cite{torquato2003local}. Such suppression of long range density fluctuations has been shown to facilitate the appearance of large PBG. The importance of uniform local topology in addition to long range hyperuniformity has been underlined in multiple publications \cite{florescu2009designer,froufe2016role}. This leads to the local self-uniformity concept, proposed by Florescu et al. \cite{sellers2017local}, which states that the best disordered PBGs correspond to networks with similar local elements, such as individual nodes,  and proposed mathematical tools to estimate the degree of similarity. Such identical elements play a role of identical Mie scatters opening a PBG by superposition of their Mie resonances \cite{rockstuhl2006correlation, chern2008optimal, edagawa2014photonic}. In the meantime,  strong local order of such networks decreases long range density fluctuations and confers hyperuniformity to the whole system, thus explaining the early observed correlation between hyperuniformity and PBG \cite{froufe2016role}.

So far, the fabrication of disordered PBG materials is dominated by the top-down approach, such as photolithography or 3D printing \cite{man2013photonic,man2013isotropic,muller2014silicon}, for which serious bottlenecks are slowness, cost and, in some cases, insufficient resolution. This is probably related to the fact that most of the proposed design techniques are based on mathematical algorithms and do not give a direct self-assembly compatible fabrication method.

In the meantime, a number of self-assembling four fold networks have been realized and  studied for years by soft matter scientists, such as dry foam/emulsion structures \cite{weaire2001physics}. The local structure of  relatively dry foams and emulsions is constrained by Plateau laws: the dry foam structure can be represented as a network of slender channels called Plateau borders which meet at vertices called nodes. Every node connects four Plateau borders (three in 2D) such that any angle between any two adjacent Plateau borders is $109.47^\circ$ in 3D (tetrahedral angle) and $120^\circ$ in 2D \cite{weaire2001physics, cantat2013foams}. Foams also tend to equilibrate the length of Plateau borders  and avoid very long and very short channels \cite{saye2013multiscale, kraynik2003foam, cantat2013foams}. At constant volume fraction, the pressure equilibrium between bubbles also gives rise to a constant thickness of the Plateau borders \cite{cantat2013foams}. This gives foam a very strong local order, very similar to local self-uniformity concept, which according to \cite{sellers2017local} is crucially important to open PBG. Moreover, these structural properties are self-sustaining: in case of any destruction event (for example, coalescence of two bubbles), the foam will immediately rearrange its structure to again fullfill the Plateau laws. Foam has been shown to be an interesting material possessing a phononic band gap for acoustic waves \cite{pierre2014resonant}. The described properties, such as strong similarity of foam structure units and robustness of its local organization, as well as the overall resemblance of dry foam structure to amorphous diamond allow us to speculate that, having the appropriate refractive index contrast between "liquid" and "air" phases, foams could produce a large robust PBG for wavelengths comparable with the bubble size.

To check this hypothesis,  we report here the use of two-dimensional foams (monolayers of bubbles squeezed between two plates)  as potential templates to generate self-assembled 2D material with a transverse electric (TE, electric field in the plane) omnidirectionnal PBG. Foams are both created experimentally and simulated numerically. Their photonic properties are calculated by a plane wave expansion method. The structural organization of relatively dry 2D foams (such as 3-fold connectivity of Plateau borders, for example) turns out to be especially useful to open a large PBG comparable with the best ones described in literature for optimized disordered materials. Our results show that the small polydispersity of bubble sizes introduce a controlled level of disorder to foam structure which  makes the band gap isotropic without strongly perturbing photonic performance. Such two-dimensional foams are well known to capture the main properties of real 3D foams, so it provides strong evidence that our results in 2D can be directly transferred to real 3D foams.

We believe that this research not only can define a roadmap to a new class of self assembled photonic materials but also sheds a light on the fundamental questions of PBG physics.

\section{Protocol}

To produce two-dimensional foams a home-made Hele-Shaw cell consisting of two vertical glass plates separated with a gap of 1.5 mm is used. Two populations of bubbles are generated by blowing air with traces of $C_{6}F_{14}$ through two orifices into a sodium dodecyl sulfate solution. The surfactant concentration is kept constant at 12 g/L (approximately 5 times the critical micelle concentration) to avoid any surfactant depletion during the
generation of the foam. The diameter of the orifices is 0.3 mm leading to bubbles of 2-3 mm in diameter. By controlling the air pressures at both entrances independently, we generate bidisperse foams with well-controlled  size and number ratios. Patterns inside the Hele-Shaw cell made with PDMS  ensure the mixing between the two populations. To get a foam with homogeneous liquid area fraction $\phi$ the experiments are performed in the forced drainage regime: foaming liquid is continuously added from the top of the foam at a controlled flow rate. This allows us to span a very large range of foam structures, from dry foam with long and slender Plateau borders and polygonal bubbles to wet foam with more roundish bubbles, ending with bubbly liquids. To get clear images of Plateau borders, the photos are taken through a right-angle prism  glued outside the container wall as first proposed by Garrett et al.
(cited in \cite{mukherjee1995}). We ensure isotropy and absence of crystallization in every experimental image by checking the absence of Bragg peaks in its Fourier transform. 
\begin{figure}[h!]
\includegraphics[width=1\linewidth]{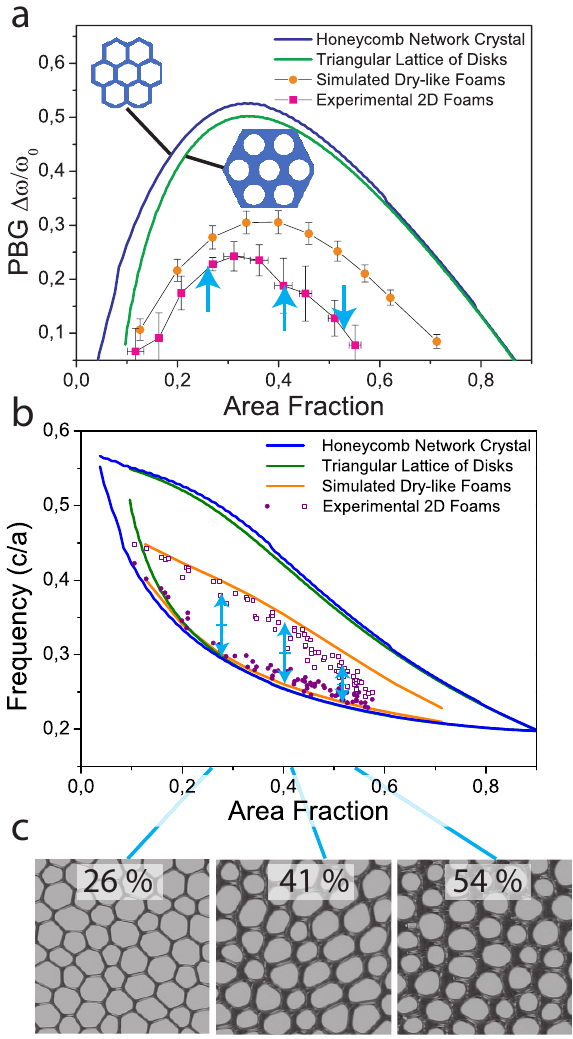}
\caption{\textbf{(a)} PBG width for TE polarization as a function of the dielectrical material area fraction for different structures: honeycomb network crystal (blue line), triangular lattice of disks (green line), simulated dry-like foam (orange circles) and experimental 2D foam (purple squares).  \textbf{(b)} Lower and upper frequencies for the Transverse Electric PBG. Blue line: lower and upper frequencies of PBG for honeycomb network crystal, green line: lower and upper frequencies of PBG for triangular lattice of disks, orange line: lower and upper frequencies of PBG for simulated dry-like foams, Purple circles and squares: lower and upper frequencies of PBG for experimental 2D foams. \textbf{(c)} Photographs of the experimental 2D foam for different area fractions: Left 26 \%, Middle 41\%, Right 54\%. Bandgaps of the corresponding systems are indicated on \textbf{(a)} and \textbf{(b)} by blue arrows.}
\label{figure2}
\end{figure}

To model  bidisperse foams in the very wet limit close to the jamming point, we generate bidisperse 2D jammed disk assemblies with periodic boundary conditions using freely available code based on Lubachevsky-Stillinger algorithm \cite{skoge2006packing,atkinson2016static}. In our previous work we have shown that this algorithm closely matches the structure of bidisperse bubbles/droplets assemblies close to jamming point \cite{ricouvier2017optimizing}. Jammed disk assemblies fabricated experimentally and  produced by the LS algorithm are shown to have a high level of effective hyperuniformity \cite{ricouvier2017optimizing, zachary2011hyperuniform1,zachary2011hyperuniform2,zachary2011hyperuniform3,dreyfus2015diagnosing}. Therefore, they are particularly interesting as potential precursors for the  novel PBG materials. By slightly decreasing the radii of disks we can closely mimic the bubbly liquid structure observed experimentally.

To get samples of foams with periodic boundary conditions in the dry limit, weighted Voronoi tessellation is performed over the obtained disordered disk assemblies. In order to bring our Voronoi tessellation, which does not necessary strongly respect Plateau's laws, closer to a real foam, we anneal the structure with Surface Evolver \cite{Brakke1992} minimizing the total perimeter of the network under the constraint of a constant area of individual bubbles. The Surface Evolver optimization of the obtained structure ensures the fulfillment of the Plateau laws so that the lengths and the angles of Plateau borders are uniform, thus enlarging the PBG (see SM - Figure S1). By increasing the thickness of the Plateau borders, we get periodic network structures with a controlled, variable quantity of dielectric material still keeping topology of dry foam. According to ref \cite{tong2017geometry}, these steps ensure that the simulated foam is very close to experimentally generated foams in a dry limit: the simulated dry-like foams always respect Plateau laws by keeping coordination 3 and angle 120 between Plateau borders. For higher area fractions, where bubbles get roundish, such "pseudo dry" foam structures do not represent a real foam anymore, but still stay interesting for PBG investigation.

\begin{figure*}[h!]
\centering
\includegraphics[width=1\linewidth]{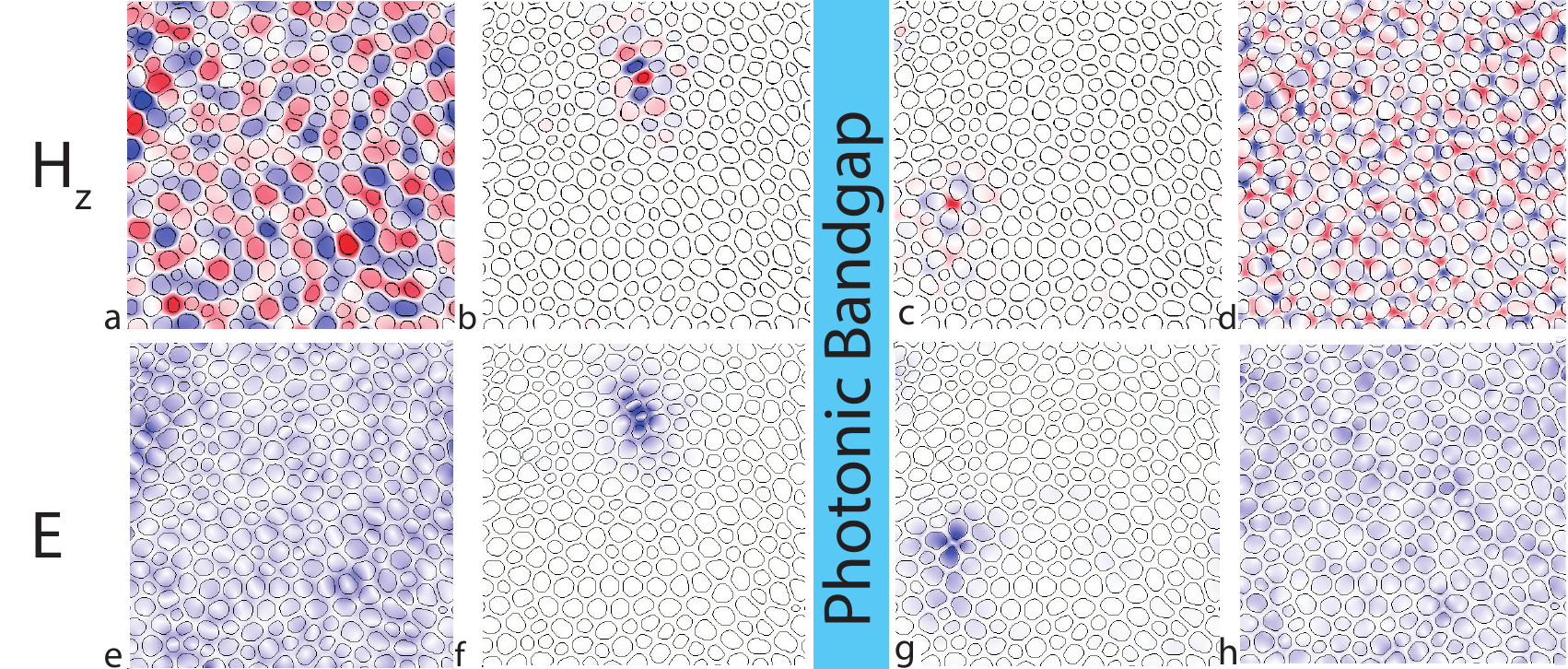}
\caption{\textbf{(Top)} Azimuthal magnetic field distribution in an experimental 2D foam for TE polarization, fraction of dielectric material is $44\%$. (a) Extended mode before the PBG. (b) Localized mode before the PBG. (c) Localized mode after the PBG. (d) Extended mode after the PBG. Notice that the localized modes present a four-fold and a five-fold "symmetry". \textbf{(Bottom)} (e)-(h) Electric field magnitude for TE polarization for the same modes and the same system. Please note that the electric field belongs to the plane for TE polarization and therefore the magnetic field is along the z axis.}
\label{Modes}
\end{figure*}

We set the refractive index of the dispersed "air" phase at $n_{d}=1$ and the refractive index of the continuous "liquid" phase at $n_{c}=3.4$ corresponding to the refractive index of amorphous silicon  which is a common standard material in PBG calculations. Guidelines for PBG optimizations for some other refractive indices can be found in SM - Figure S2). We use the eigenstate solver MIT Photonic Bands (MPB) to compute the photonic band structure, employing the supercell approximation \cite{johnson2001block}. These calculations assume periodic boundary conditions  which, of course, cannot be achieved in experimental systems. In the case of experimental 2D foams this results in the formation of eigenstates strongly localized at the boundaries of the supercell. These eigenstates are related only to finite size effects and are not relevant to define the PBG of the bulk material. To filter these localized bands we remove all modes storing more than 95 $\%$ of energy within the layer of two bubbles close to the border of the supercell. Some examples of these boundary states are shown in Figure S3 in supplementary materials. For  simulated dry-like foams in periodic supercells, there is no need to filter eigenstates. To obtain a normalized density of optical states (NDOS) the total density of states is divided by the corresponding one of the homogeneous media with the same effective refractive index obtained from the first band of the system under investigation. For experimental 2D foams photonic bandgap calculations for each area fraction were performed on 5 to 10 images containing about 300 bubbles. For each area fraction, 10 images of disordered disk assemblies and simulated dry-like foams containing exactly 300 bubbles were generated. We set the resolution to 512x512.

\section{Results and discussion}

Figure \ref{Imagesoffoam}b shows a photograph of bidisperse two-dimensional foam produced in a Hele-Shaw cell. Size ratio of big to small bubbles, defined as the square root of corresponding Voronoi cell's areas, is about 0.75. This relatively low polyspersity  is chosen to avoid crystallisation without strongly perturbing the self-similarity of foam elements (such as length of Plateau borders). A circular symmetry of the Fourier transform and absence of Bragg peaks (Inset of Figure \ref{Imagesoffoam}b) prove that we manage to create a disordered isotropic foam without spatially extended crystal domains. By virtually replacing the liquid phase with silicon in our calculations ($n=3.4$), we observe that such disordered foams demonstrate an isotropic PBG which does not depend on the direction of propagation: the modes close to the band gap are almost flat. We find that the bandgap width $\Delta\omega /\omega_0$, defined as a ratio of band gap size $\Delta \omega$ to middle frequency $\omega_0$, reaches $~25 \%$ for experimental 2D foam which is comparable to the best disordered systems described in literature ($~30 \%$ for TE polarization in 2D) \cite{florescu2009designer}. This makes 2D foam one of the first and most promising examples of self-assembled disordered PBG materials.

Figures \ref{Imagesoffoam}a and c show two reference simulated systems which represent two extreme cases: dry-like foam obtained with Surface Evolver (simulated dry-like foam) and bidisperse disk assembly (disordered disk assembly) representing foam in a very wet limit. The consideration of these two simulated systems allows to study the effect of foam structure separately from the dielectric material fraction influence. One can see that for the same dielectric material area fraction, the simulated dry-like foam has a large band gap comparable to the experimental one. At the same time, the wet foam system has only a pseudo-gap with strongly diminished NDOS but no sign of true PBG (see Figure \ref{Imagesoffoam}d).

To see how real foam PBG response evolves from dry to wet limit, we plot in Figure \ref{figure2} the high dielectric material area fraction $\phi$ dependence of PBG width $\Delta\omega /\omega_0$. For low quantity of dielectric material the width of the PBG $\Delta\omega /\omega_0$ for experimental 2D foam increases with the area fraction. It passes through a maximum at $\phi = 0.3 - 0.4$ and starts to decrease. Up to this point the behavior is qualitatively similar to the one observed in the corresponding honeycomb network or triangular disk crystals both having a maximum approximately at the same fraction (see Figure \ref{figure2}). The PBG for the experimental system is also quantitatively close to the one calculated for simulated dry-like foams.  Surprisingly, at about $\phi =0.45 - 0.55$ the band gap of the experimental 2D foam  abruptly closes and no PBG can be observed for higher area fractions. Foam starts to approach the wet limit, at least from the PBG point of view.  Our simulations with disordered disk assemblies never show any sizeable bandgap for any area fraction. In the case of simulated dry-like foam and crystal structures the bandgap persists for area fractions as high as $0.8$. We can conclude that dry foam structure with well defined nodes and Plateau borders gives a large PBG  at appropriate dielectric material fraction while the wet one is completely incompatible with PBG.

A closer look on the PBG frequencies reveals that in all considered systems, both monodisperse crystallized and slightly polydisperse disordered ones, the PBG (NDOS=0) or pseudo-gap (reduced NDOS) starts roughly at the same frequency for a given area fraction (see Figure \ref{figure2}). In the meantime, the upper edge of the band gap turns out to be different for various systems and thus is responsible for the change of PBG width between various systems. 

 \begin{figure*}[h!]
\centering
\includegraphics[width=1\linewidth]{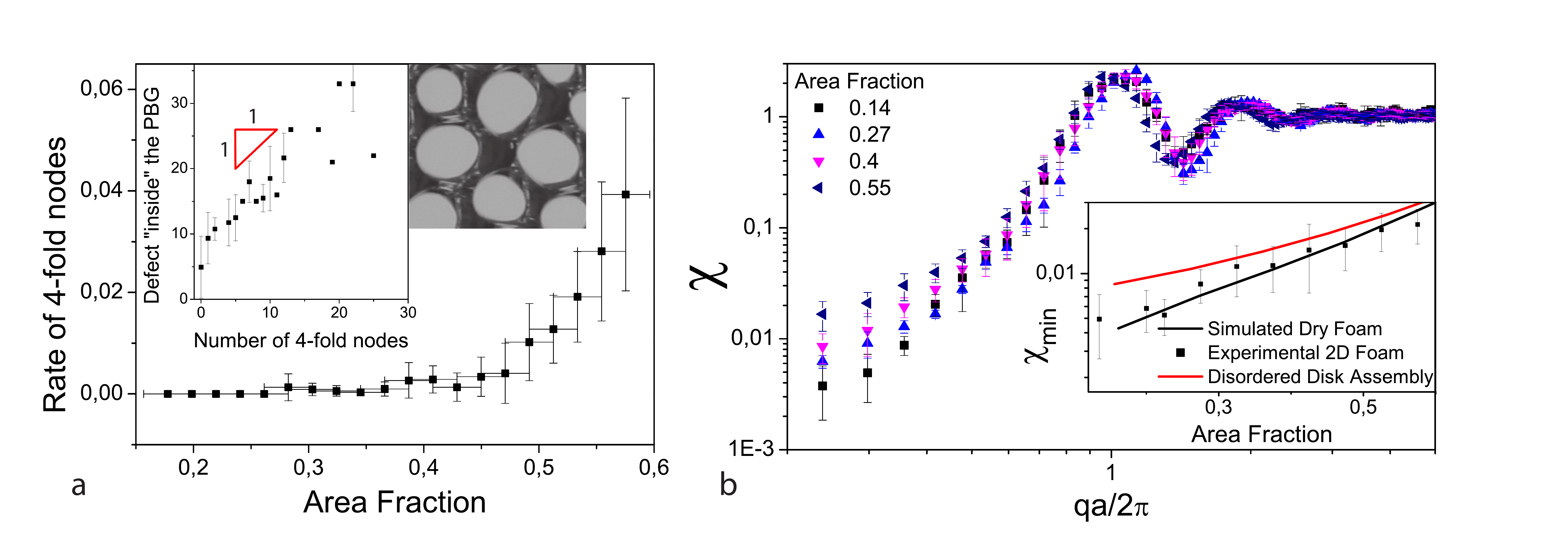}
\caption{(a) Rate of 4-fold nodes plotted along the area fraction of dielectric material for experimental 2D foams. Insets: number of defect modes in the PBG vs the number of 4-fold nodes in experimental 2D foams; a photograph of a typical four-fold node. (b)Spectral density of experimental 2D foams for several area fractions. Inset: Spectral density at minimal wave vector for different systems: disordered disk assembly (red line),  simulated dry-like foams (black line) and experimental 2D foam (black squares).}
\label{Spectraldensity}
\end{figure*}

 As shown in Figure \ref{Modes} the dielectric bands just below the PBG  are strongly localized in individual bubbles for the magnetic field and in the dielectric material around the bubbles for the electric field as previously shown for similar network structures \cite{florescu2009designer}. This result stays valid even for the disordered disk assemblies, which have a very narrow range of frequencies, where NDOS approaches zero. Clearly the lower edge frequency depends mainly on the mean bubble size and area fraction, also influencing the separation between the bubbles. It does not strongly depend on the exact bubble shape (polygons or disks), nor on the structural organization of the system (crystallized or not). Passing through the bandgap both fields change their localization as is well-known both for crystalline and disordered PBG materials \cite{joannopoulos2011photonic, florescu2009designer}. The air bands at the upper edge of the bandgap are rather localized around individual foam nodes. The magnetic field is localized inside the nodes while the electric field is concentrated in bubbles adjacent to the nodes.

We should naturally expect that the change of the PBG upper edge for different systems, such as a rapid closing for wet foams in comparison with the dry ones, should be related to the change of the node organization. Indeed we observe that for relatively wet 2D foams (area fraction above $0.4$), where the PBG width $\Delta\omega /\omega_0$ starts to diminish, the first modes after PBG correspond to the 4-fold nodes as shown in Figure \ref{Modes}. Such 4-fold nodes clearly do not respect Plateau laws and can appear only when foam goes to the wet limit. To prove that such nodes are indeed responsible for the PBG suppression for wet foams we plot a fraction of such nodes as a function of area fraction as shown in Figure \ref{Spectraldensity}. One can see that these defect nodes appear for area fraction higher than $0.4$. The amount of 4-fold nodes rapidly explodes with increasing dielectric material quantity in the same range of area fractions where the PBG abruptly closes down. We recall that our simulated dry-like foams obey the Plateau laws even for relatively high area fractions  and do not have such 4-fold nodes by construction. The rate of 4-fold nodes can also serve to define the limit (about 35-40 \% of area fraction) where experimental foam cannot be represented by simulated dry-like foam.

If we compare experimental 2D foams and simulated dry-like foams with the same area fraction, electromagnetic modes related to these 4-fold nodes can be considered defect modes in the PBG of idealized dry foams. The 4-fold nodes have a lower characteristic frequency than 3-fold nodes, related to their bigger characteristic size. So the 4-fold nodes create defect states that start to fill the PBG of dry foam from the upper edge and gradually diminish it with increase of area fraction. To highlight this strong correlation between PBG width and percentage of 4-fold nodes, we show for experimental 2D foams in the Inset of Figure \ref{Spectraldensity} the number of defect modes as a function of 4-fold nodes number. The number of modes inside the PBG clearly scales linearly with the number of 4-fold nodes throughout all our data. This gives a straightforward way to estimate the PBG performance of the foams, simply by considering their topology. This linear scaling also means that if  if we removed the states corresponding to these 4-fold nodes, the PBG of experimental foams would follow the PBG of simulated dry-like foams in the whole range of studied area fractions. This, for example, reveals that geometric factors, such as local curvature of nodes or circularity of bubbles (see Figure S4 for circularity of experimental and simulated dry-like foams), is not particularly important for the PBG in TE polarization. This phenomenon can also be seen for the crystalline structures, where both honeycomb networks and triangular lattice of disks have almost identical PBG for TE (See Figure \ref{figure2} and Figure S2 in SM). 

The analogy between the PBG of crystallized materials and our systems stays valid for TM polarization. No sizeable TM PBG is found in honeycomb network lattice  as well as in our systems (see Figure S5).  Similar  networks of walls described in literature possess large PBG for TE polarization, but no band gap for TM. However, this 2D network has given rise to 3D PBG structures, such as amorphous diamond, underlining a particular importance of TE polarization.

Summing up, we can conclude that the lower frequency of the TE PBG of the studied systems  is mainly related to the fraction of the dielectric material and the size of the bubbles, no matter how they are organized (crystallised or not) or the shapes the bubbles adopt (disks or polygons). The upper frequency is related to several phenomena, the most important one is the diversity of node shapes: the more diverse the nodes are, the smaller the PBG is. This approach also explains why disordered disk assemblies, representing very wet foams at jamming point, never show a sizeable PBG. Nodes in such system can have various shapes, even more complicated than the 4 fold nodes presented in Figure \ref{Modes}, and they create defect states that completely fill the PBG.

The appearance of a band gap is often related with the degree of hyperuniformity in the system, so it is also interesting to evaluate the hyperuniformity of our foams. Figure \ref{Spectraldensity} shows the spectral density  of the foam which is an extension of structure factor for bidisperse and generally polydisperse systems \cite{wu2015search,berthier2011suppressed}.  In the inset, we also plot the spectral density at the minimum wave vector $\chi_{min}$ which has been proven to be a convenient way to estimate the hyperuniformity of experimental systems \cite{ricouvier2017optimizing}. The minimum spectral density $\chi_{min}$ increases with the area fraction  for all considered systems.  One can see that for a given area fraction  both experimental 2D foams and simulated dry-like foam structures systematically have less long range density fluctuations (lower $\chi_{min}$) than disordered disk assemblies which have no PBG. According to previous works \cite{dreyfus2015diagnosing,berthier2011suppressed,donev2005unexpected}, such disk assemblies at jamming point (area fraction around 16 \%), are accepted to be hyperuniform. Our results show that simulated dry-like and experimental 2D foams provide an even higher level of hyperuniformity (lower $\chi_{min}$) while possessing a large PBG. Thus, at fixed area fraction, the decrease of $\chi_{min}$ for simulated dry-like and experimental foams, in comparison to disk assemblies, seems to be correlated with the PBG opening.  This result is in agreement with the previously observed correlation between PBG and hyperuniformity \cite{florescu2009designer,froufe2016role}.

In the meantime, the spectral density is not sensitive enough to capture the closing of the PBG for experimental foams going from the dry to the wet limit. We cannot find any special transition neither for the spectral density  $\chi (q)$  nor for the minimum spectral density $\chi_{min}$ about the area fraction  $0.4$. Minimal spectral densities of experimental 2D and simulated dry-like foams are always similar within the error bars.  This is why we believe that local order, represented, for example, by the rate of the 4 fold nodes, is a more sensitive tool to assess the PBG performance. Neither spectral density nor $\chi_{min}$ is sensitive enough to catch the appearance of 4-fold nodes. As discussed above, the concentration of the local defects explains the difference between all studied systems: absence of PBG for disk assemblies, as well as the degradation of the PBG at the transition of dry to wet experimental foams. This result supports once more that the large and robust PBG observed in our photonic foams is not only due to hyperuniformity but is also related to the local organization of the system, such as node coordination, for example.

With the results obtained for real and simulated dry 2D foam structures, we can expect that dry 3D foam structures should exhibit PBG as well. We can give a rough estimate of expected PBG width for 3D foams from the results published for structurally similar amorphous diamond networks of dielectric rods.  Amorphous diamonds have been numerically optimised to possess a large isotropic PBG of around 20 \% for a volume fraction of dielectric material of around 20-30\% and the refractive index contrast $3.4$ \cite{edagawa2008photonic}. This volume fraction of dielectric material is beyond the accepted limit of dry foams in 3D, which is rather about 5-10\% \cite{cantat2013foams}.  This would mean that to produce a good candidate for complete PBG 3D photonic foam by self-assembly, first a dry solid foam from high refractive index material should be fabricated and then the quantity of dielectric material should be additionally  increased by coating the Plateau borders. A high reliability to disorder of two-dimensional dry foams also suggests that the PBG of proposed three dimensional photonic foams should be much more robust than the PBG of widely used inverse opal crystals: inverse opals are structurally close to wet foams, any disorder creates a large variety of nodes and rapidly closes the PBG. This makes 3D photonic foams a promising candidate for a next generation of self-assembled complete isotropic PBG materials.

\section{Conclusion}

We have demonstrated that two-dimensional foams can be used as a self-assembling template for disordered materials with a large isotropic PBG for TE polarization. We show that the size and position of the band gap can be changed by adjusting the fraction of dielectric material opening a way for tunable bandgap applications. The dry foam structure turns out to be especially profitable to open the PBG - high level of hyperuniformity at the relevant lengthscale and significant similarity of dry foam nodes (for example, the 3 fold coordination in 2D) guarantees a large and isotropic PBG to photonic foams. However, the photonic performance strongly diminishes in the wet limit due to appearance of defect states in large four-fold nodes. Disordered disk assemblies, mimicking the structure of bubbly liquids, never show a sizable PBG. This can be correlated both with the increase of long range density fluctuations and the loss of a uniform local topology. The described 2D photonic foams can already be used as a platform to fabricate planar optical circuits. However, we also expect our results to stay valid for real 3D foams being four fold network of channels. Our research suggests that photonic foams are potentially an excellent candidate for  materials with a large complete isotropic photonic bandgap. Photonic foams are fully scalable to a wide range of wavelengths from visible range to microwaves and potentially can be produced in a large quantity by self-assembly.

We are much grateful to K. Morozov, A. Leshansky, N. Stern, L. Froufe-P\'erez, F. Scheffold and A. Salonen for fruitful discussions and suggestions made along the work. This work has been supported by ESPCI Paris, PSL and IPGG. The Microflusa project receives funding from the European Union Horizon 2020 research and innovation programme under Grant Agreement No. 664823.
% Display the acknowledgments section

%\showacknow{}  % Display the acknowledgments section

\end{document}